**The role of post-shock heating by plastic deformation during impact devolatilization of calcite**

Kosuke Kurosawa[1,*], Hidenori Genda[2], Shintaro Azuma[3], and Keishi Okazaki[4]

[1]Planetary Exploration Research Center, Chiba Institute of Technology, 2-17-1, Narashino, Tsudanuma, Chiba 275-0016, Japan.

[2]Earth–Life Science Institute, Tokyo Institute of Technology, 2-12-1 Ookayama, Meguro-ku, Tokyo 152-8550, Japan.

[3]Department of Earth and Planetary Sciences, Tokyo Institute of Technology, 2-12-1 Ookayama, Meguro-ku, Tokyo 152-8550, Japan.

[4]Kochi Institute for Core Sample Research, Japan Agency for Marine-Earth Science and Technology, 200 Monobe Otsu, Nankoku, Kochi 783-8502, Japan.

*Corresponding author
Kosuke Kurosawa PhD
Planetary Exploration Research Center, Chiba Institute of Technology
E-mail: kosuke.kurosawa@perc.it-chiba.ac.jp
Tel: +81-47-4782-0320
Fax: +81-47-4782-0372

**Abstract**

An accurate understanding of the relationship between the impact conditions and the degree of shock-induced thermal metamorphism in meteorites allows the impact environment in the early Solar System to be understood. A recent hydrocode has revealed that impact heating is much higher than previously thought. This is because plastic deformation of the shocked rocks causes further heating during decompression, which is termed post-shock heating. Here we compare impact simulations with laboratory experiments on the impact devolatilization of calcite to investigate whether the post-shock heating is also significant in natural samples. We calculated the mass of $CO_2$ produced from the calcite, based on thermodynamics. We found that iSALE can reproduce the devolatilization behavior for rocks with the strength of calcite. In contrast, the calculated masses of $CO_2$ at lower rock strengths are systematically smaller than the experimental values. Our results require a reassessment of the interpretation of thermal metamorphism in meteorites.

1. **Introduction**

Mutual collisions between two small bodies in the main asteroid belt region of the Solar System at velocities higher than several km s$^{-1}$ cause heating of their surface materials [e.g., Ahrens and O'Keefe, 1972]. This impact heating results in thermal metamorphism [e.g., Stöffler et al., 2018], including devolatilization of volatile-bearing minerals [e.g., Kieffer and Simond, 1980; Kurosawa et al., 2012], changes in optical properties [e.g., Hiroi et al., 1996], resetting of $^{40}Ar/^{39}Ar$ ages [e.g., Cohen, 2013], and melting [e.g., Ahrens and O'Keefe, 1972]. It is widely accepted that the degree of impact heating depends strongly on the impact velocity [e.g., Ahrens and O'Keefe, 1972]. Given that we have an accurate understanding of the relationship between the degree of thermal metamorphism and impact conditions, thermal metamorphic features allow us to investigate the impact environment in the Solar System [e.g., Marchi et al., 2013].

The effects of the material strength of rocky materials on impact processes have been mainly studied from a mechanical viewpoint, such as the morphology and damage structure of impact craters [e.g., Ivanov et al., 1997; Collins et al., 2004; Goldin et al., 2006; Elbeshausen et al., 2009]. Recently, the role of rock strength in terms of thermodynamics have been investigated [Quintana et al., 2015; Kurosawa and Genda, 2018]. Plastic deformation of shocked rocks working against a finite yield strength efficiently converts kinetic energy in the impact-driven flow field into internal energy in the shocked rocks, particularly during low-speed collisions (<10 km s$^{-1}$), resulting in post-shock heating. As a result, the degree of impact heating during low-speed collisions is much higher than previously expected. Post-shock heating due to plastic deformation is less important during high-speed collisions (>10 km s$^{-1}$) compared with low-speed collisions because the plastic work near the impact point is limited by thermal softening [e.g., Pierazzo and Melosh, 2000] (see the Supplementary Information).

Although post-shock heating has been reported in several studies [Ivanov et al., 2002; Kenkmann et al., 2013; Quintana et al., 2015], Melosh and Ivanov (2018) noted "this heat source has surprisingly escaped explicit attention for decades". This could be due to the difference in geometries between natural impacts and conventional uniaxial shock compression experiments [Melosh and Ivanov, 2018]. If the temperature increase due to plastic deformation is also significant for natural impacts, then the impact environment inferred for the early Solar System from meteorites would be less dynamic than previously thought, because low-speed collisions might cause significant thermal metamorphism. Post-shock heating has been explored mainly using shock physics codes, which are hydrocodes that can model the elasto-plastic behavior of rocky materials. However, validation of the numerical model by experiments on natural rocky materials remains to be conducted.

In this study, we explored whether the numerically determined heat source is physically real by comparing the calculated results with previously published experimental data on the impact devolatilization of natural calcite. Kurosawa et al. (2012) measured the mass of $CO_2$ ($M_{CO2,exp}$) from calcite marble in an open system, which was in the same geometry as natural impacts (see the Supplementary Information for details on their experiments). The mechanical and thermodynamic properties of calcite marble have been extensively studied under both static and dynamic conditions [e.g., Kerley, 1989; Hacker and Kirby, 1993; Ivanov and Deutsch, 2002; Kurosawa et al., 2012], which enables modeling of the behavior of shocked calcite with a high accuracy. Since the $CO_2$ amount released due to impact devolatilization depends strongly on the degree of impact heating, impact devolatilization of $CO_2$ is ideal for investigating whether post-shock heating occurs due to plastic deformation in natural samples.

## 2. Numerical model

### 2.1. Numerical model description and calculation settings

The two-dimensional version of the iSALE shock physics code (iSALE-Dellen; [Amsden et al., 1980; Ivanov et al., 1997; Wünnemann et al., 2006; Collins et al., 2016]) was used. We employed cylindrical coordinates to model vertical impacts performed by Kurosawa et al. [2012]. We used the "ROCK" model in the iSALE package [e.g., Ivanov et al., 1997] to account for the elasto-plastic behavior of marble. This model is a combination of the Lundborg and Drucker–Prager models [Lundborg, 1968; Drucker and Prager, 1952] for intact and damaged rocks, respectively. The two models are combined with a damage parameter, which is determined by the total plastic strain. The effect of thermal softening on yield strength (*Y*) was also included, using the formulation of Ohnaka (1995) as follows:

$$Y(T) = Y_C \tanh\left[\xi\left(\frac{T_m}{T} - 1\right)\right], \quad (1)$$

where $Y_C$, $\xi$, $T_m$, and $T$ are the yield strength at the reference temperature, a constant, the temperature at the solidus, and temperature, respectively. The formulation yields $Y(T_m) = 0$ at the solidus.

The parameter set for limestone [Goldin et al., 2006] was used for the calcite targets, except for the von Mises limit $Y_{\text{lim}}$, which largely controls the pressure dependence of the yield strength. The introduction of limiting strengths into hydrocodes can account for plastic deformation [Ivanov et al., 1997]. The value of $Y_{\text{lim}}$ proposed by Goldin et al. (2006) is $Y_{\text{lim}} = 0.65$ GPa (hereafter, referred to as the fiducial value). To test the accuracy of the yield strength obtained by the ROCK model, we compared the pressure–stress curve with the data from a static deformation experiment [Hacker and Kirby 1993]. The ROCK model with the fiducial value (0.65 GPa) for $Y_{\text{lim}}$ reproduces the strength behavior of marble as shown in Figure 1a. However, Hacker and Kirby [1993] reported that the yield strength increases with increasing strain rate at strain rates from $10^{-6}$ to $10^{-4}$ s$^{-1}$, which are orders of magnitude smaller than in the laboratory impact experiment (ca. $10^6$ s$^{-1}$). This indicates that the yield strength in the impact experiments was higher than the fiducial value. By changing $Y_{\text{lim}}$ from the fiducial value, we systematically explored the role of material strength on the degree of impact heating. Figure 1b shows the pressure–stress curves at room temperature at $Y_{\text{lim}} = 0.65$ and 1 GPa, which highlights that a variety of pressure–stress curves can be obtained by changing $Y_{\text{lim}}$. To examine the dependence of material strength on the degree of impact heating, we used different values from $Y_{\text{lim}} = 0.65$ GPa to larger values up to $Y_{\text{lim}} = 2.6$ GPa.

The analytical equation of state (ANEOS) [Thompson and Lauson, 1972] for calcite [Pierazzo et al., 1998] was used for the targets. The Tillotson EOS [Tillotson, 1962] with parameters for Al$_2$O$_3$ (Supplementary Information) was used for the projectiles because Kurosawa et al. [2012] used $Al_2O_3$ spheres as projectiles. For simplicity, we did not include any strength models for the projectile because we do not focus on the projectile behavior in this study. The projectile was divided into 200 cells per projectile radius (CPPR). Hereafter, the number of CPPR is referred to as $n_{CPPR}$. The value of $n_{CPPR}$ is a good indicator of the spatial resolution. We used a four-fold $n_{CPPR}$ with respect to the Kurosawa and Genda [2018] to estimate the heated volume with a high degree of accuracy. The calcite target was set to be a cylinder with 2000 cells in the radial direction and 2600 cells in the depth direction. This target size is large enough to prevent the effects of wave reflection from the boundaries of the computational domain. Lagrangian tracer particles were inserted into each cell. We recorded the temporal pressure $P$ and entropy $S$ in each tracer. The impact velocities $v_{imp}$ in the simulation were set to the same values as in the laboratory impact experiments [Kurosawa et al., 2012]. Further details can be found in Supplementary Information.

## 2.2. $CO_2$ production in the iSALE computation

Previous numerical studies related to impact devolatilization of calcite used a method termed the "peak-pressure method" [e.g., Takada and Ahrens, 1994; Pierazzo et al., 1998; Artemieva et al., 2017] to estimate the total $CO_2$ emissions. This method assumes that the shocked calcite follows an isentropic path. The "critical pressure" on the Hugoniot curve required for incipient decomposition of calcite after isentropic decompression down to the ambient pressure is then calculated by EOS models. There are three problems in this approach to determining the impact devolatilization of calcite. The first is that the assumption of isentropic release is not valid if the post-shock heating occurs efficiently, as shown by Kurosawa and Genda [2018]. The second problem is that the ANEOS for calcite does not accurately predict the decomposition boundary, resulting in an uncertainty in the degree of decomposition of calcite. What is actually modeled by the current version of ANEOS calcite is "vaporization" of calcite into diatomic gases [e.g., Pierazzo et al., 1998; Ivanov and Deutcsh, 2002], and not "devolatilization" (i.e., $CaCO_3(s) \rightarrow CaO(s) + CO_2(g)$). Significantly higher entropy is required for complete vaporization than that for complete devolatilization. As such, it is challenging to robustly estimate the degree of decomposition (or the efficiency of the back reaction). The third problem is that experimental results estimating the critical pressure required for incipient devolatilization range from 20 to 60 GPa [e.g., Boslough et al., 1982; Ohno et al., 2008;

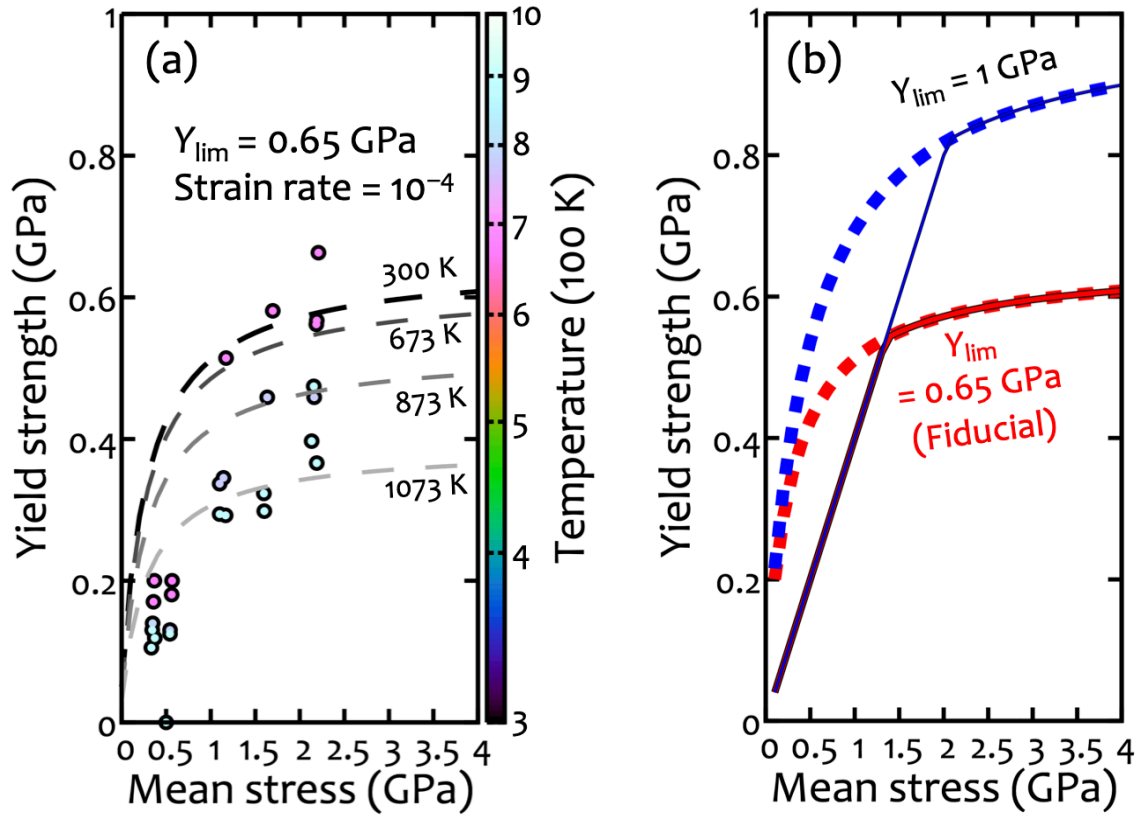

**Figure. 1.** Pressure- and temperature-dependent variations in yield strength. (a) The pressure- and temperature-dependence of the yield strength. The dashed lines are the pressure–stress curves at given temperatures. The temperature is annotated next to each line. The circles show the experimental data obtained at a strain rate = $10^{-4}$ s$^{-1}$ [Hacker and Kirby, 1993], and the circle colors indicate the temperature conditions of the experiments. (b) The effects of the von Mises limit on the pressure–stress curve. The solid straight line corresponds to the yield strength of the damaged calcite. The yield strength pertaining to intact calcite ($Y_i$; dashed line) limits the yield strength for damaged calcite ($Y_d$; solid line). The red dashed curve is the same as the line for 300 K plotted in (a).

Kurosawa et al., 2012; Bell, 2016]. Thus, previous modeling studies selected a single value of critical pressure for decomposition from the experimental values. The scatter in these data results in an uncertainty in the total $CO_2$ emissions within a factor of 2–3 [Artemieva et al., 2017]. It is noted that another method, termed the "final-temperature method", is available to estimate heated masses during impacts [Quintana et al., 2015], although the method has not yet been applied to impact devolatilization of calcite. The final-temperature method, however, does not provide the degree of devolatilization in
6

principle. Although the region of partial decomposition degenerates into a curve on a pressure–temperature plane, the degree of decomposition depends on entropy even at the same pressure and temperature.

We employed a two-step method to overcome the three problems as follows. Figure 2 shows a phase diagram for calcite on an entropy-pressure plane. We computed the change in entropy of calcite due to shock loading and pressure release down to 1 GPa with the iSALE. The effects of the post-shock heating due to plastic deformation are also taken into account in this calculation. The post-shock heating does not occur at pressures below 1 GPa [Kurosawa and Genda, 2018]. We then assumed isentropic decompression of shocked calcite. The assumption, which is that the shocked calcite adiabatically expands from 1 GPa to the ambient pressure $P_{amb}$, is justified because the cooling rate due to adiabatic expansion is much higher than that due to other cooling processes [e.g., Sugita and Schultz, 2002]. We calculated the degree of devolatilization $\psi$ of each tracer particle using the lever rule and by assuming that the system was in thermal equilibrium as follows:

$$\psi = \begin{cases} 0, & (S < S_{id}) \\ \frac{(S-S_{id})}{(S_{cd}-S_{id})}, & (S_{id} < S < S_{cd}), \\ 1, & (S_{cd} < S) \end{cases} \quad (2)$$

where $S_{id}$ and $S_{cd}$ are the entropies required for incipient and complete devolatilization, respectively. The decomposition boundaries of calcite ($S_{id}$ and $S_{cd}$) were determined by Kurosawa et al. [2012] using the NASA CEA code [Gordon and McBride, 1994]. The thermochemical data for related species in the Ca–C–O system have been extensively explored [e.g., Chase, 1985]. The location of the decomposition boundary of calcite as a function of entropy could be accurately estimated below the pressure where the ideal gas assumption for the produced $CO_2$ is valid (i.e., lower than the atmospheric pressure). The entropies $S_{id}$ and $S_{cd}$ at $P_{amb}$ are estimated to be 2.2 and 3.9 kJ K$^{-1}$ kg$^{-1}$, respectively, in the case of $P_{amb}$ = 2.7 kPa. In the peak pressure method, which is frequently used in the previous studies as mentioned above, $S_{id}$ and $S_{cd}$ correspond to 47 GPa and 169 GPa, respectively [Kurosawa et al., 2012]. The expected $CO_2$ production $M_{CO2}$ after pressure release was estimated to be:

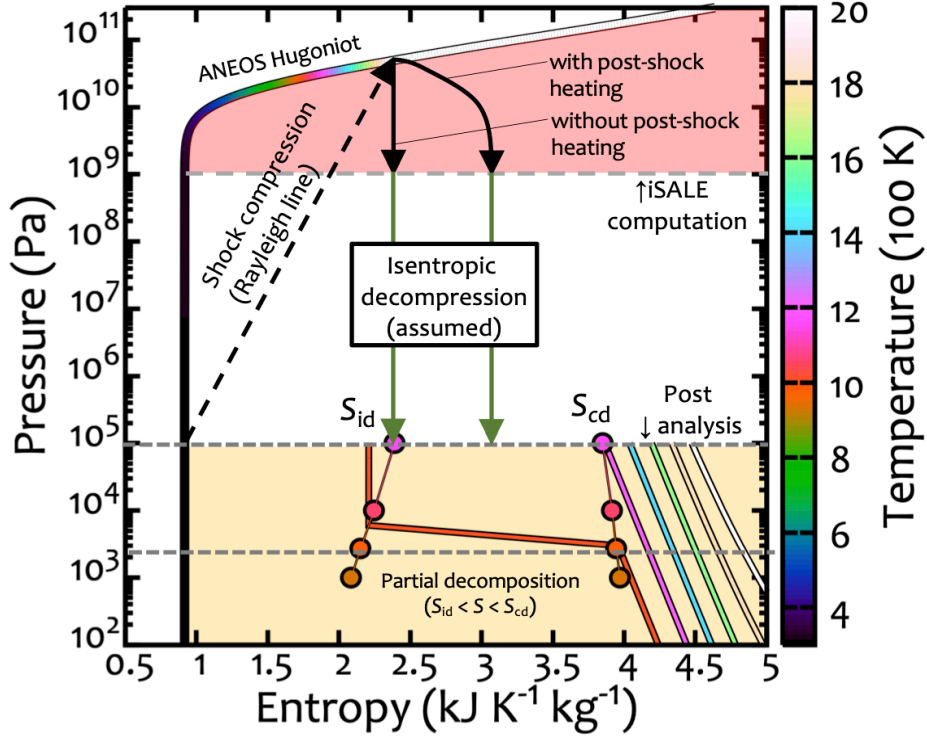

**Figure 2.** Phase diagram of calcite on an entropy–pressure plane. The Hugoniot curve for calcite by ANEOS is shown. The red shaded region enclosed by the Hugoniot curve and horizontal line at 1 GPa was determined by iSALE. The yellow shaded region below $10^5$ Pa was determined based on thermodynamics. The phase boundaries for both incipient and complete decomposition calculated by the CEA code are shown as circles. We assumed isentropic decompression below 1 GPa (see text). The schematic changes in entropy in the cases with and without post-shock heating are shown as black arrows. The dashed horizontal line at 2.7 kPa corresponds to the reference pressures for the laboratory experiment by Kurosawa et al. [2012]. For reference purposes, the isotherms of the Ca–C–O system calculated with the CEA code from 1,000 to 2,000 K, at steps of 200 K, are also shown. The color bar shows temperature.

$M_{CO2} = \Sigma(0.44\,\psi_i \Delta m_i)$, (3)

where the subscript $i$ indicates the tracer ID and $\Delta m$ is the mass of the tracer particle in cylindrical coordinates. The constant 0.44 corresponds to the ratio of molecular weights of $CO_2$ to $CaCO_3$. We used the $S$ value at 1 GPa stored on each tracer to calculate $M_{CO2}$ at each time step. To exclude weakly shocked materials (peak pressure < 3 GPa), we only used tracers that were once shocked to >3 GPa, but were released down to 1 GPa. As such, we did not use the value after decompression down to $P_{amb}$. This approach largely prevents numerical diffusion and integration error. We confirmed that $M_{CO2}$ became constant in all cases by $t = 5.1\ t_s$ in the simulations. The characteristic time $t_s$ is defined to be $t_s = 2R_p/v_{imp}$, where $R_p$ is the projectile radius. Consequently, we used the $M_{CO2}$ value at the time after $M_{CO2}$ became constant as $M_{CO2,calc}$, and then used this to compare with $M_{CO2,exp}$.

## 3. Results

Figure 3a shows a snapshot of the calculations with $Y_{lim} = 0.65$ GPa and $v_{imp} = 4.14$ km s$^{-1}$ along the trajectories of five selected tracers initially located at $r = 0.5R_p$, where $r$ is the radial distance from the symmetry axis. The radial distance $r = 0.5R_p$ was chosen as it is inside the impactor footprint and far from the symmetry axis. The highly heated calcites are concentrated near the crater wall. Figure 3b shows the thermodynamic tracks on a $S$–$P$ plane of the tracers initially located at the same radial distance as the five selected tracers. The entropies of the five selected tracers suddenly increase immediately after shock arrivals. The entropies then further increase during decompression, which is post-shock heating due to plastic deformation. Finally, the entropy increase ceases at ca. 1 GPa. The condition where $S > S_{id}$ is achieved even at the peak pressure of 25 GPa, which is much lower than the required peak pressure for incipient devolatilization (47 GPa) estimated by the peak pressure method. For comparison, the results at the same impact velocity, but without material strength, are shown in the Supplementary Information. The entropy increase during decompression does not occur in this case.

Figure 4 shows a comparison between $M_{CO2,exp}$ and $M_{CO2,calc}$. We found that $M_{CO2,calc}$ in the hydrodynamic and $Y_{lim} < 1.3$ GPa cases is systematically lower than $M_{CO2,exp}$. In contrast, with $Y_{lim} = 1.3$–$2.6$ GPa, $M_{CO2,calc}$ is close to $M_{CO2,exp}$.

The gray hatched region in Fig. 4 indicates that $M_{CO2,calc}$ is not reliable due to the low spatial resolution. The horizontal line on the top of this region corresponds to $M_{CO2,calc}$ of 0.3% of the projectile mass. This mass is equivalent to the mass of the calcite cylinder below the impactor footprint with a thickness of one cell. It is widely known that the cumulative heated mass depends on the spatial resolution [e.g., Wünnemann et al., 2008] and, as such, we must address the spatial resolution effect on $M_{CO2,calc}$. We confirmed that $M_{CO2,calc}$ is inversely proportional to $n_{CPPR}$ under the calculation conditions of this study.

The difference in $M_{CO_2,calc}$ values between $n_{CPPR} = \infty$ and 200 is within a factor of 1.1 in most cases, although the maximum difference is a factor of 1.25 in one case. The details of the resolution test and further possible uncertainties associated with the numerical model are described in the Supplementary Information.

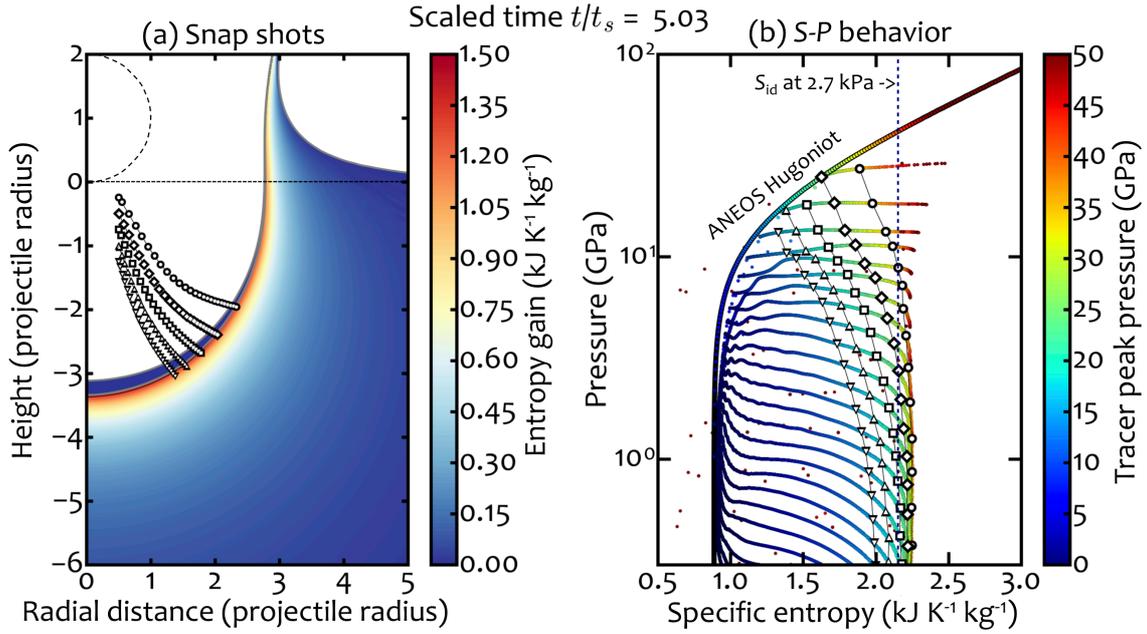

**Figure 3.** Results of the impact model. (a) Snapshot at $t = 5\ t_s$, $v_{imp} = 4.14$ km s$^{-1}$, and $Y_{lim} = 0.65$ GPa. The color indicates the entropy gain from the reference state. The trajectories of the five selected tracers (See text) are also shown. (b) The thermodynamic tracks of the five tracers on an entropy–pressure plane along the Hugoniot curve. The behavior of the tracers initially located on the same vertical line for the five tracers are also shown as colorful lines. The time interval between two lines is $0.2\ t_s$. The color indicates the peak pressures of the tracer particles. The vertical dotted line corresponds to the entropy required for incipient devolatilization at 2.7 kPa. This line also corresponds to the residual temperature of 970 K after pressure release.

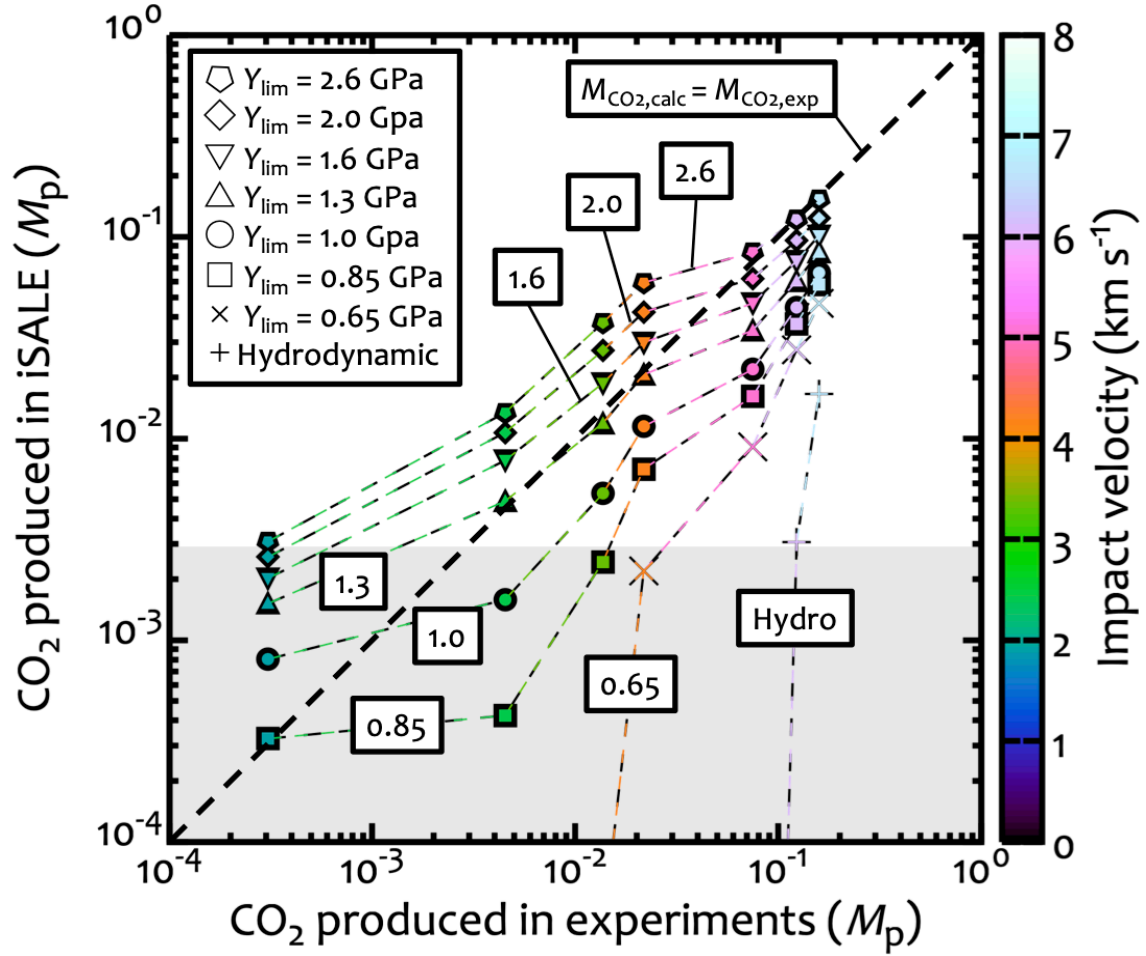

**Figure 4.** Comparison of $M_{CO2,exp}$ and $M_{CO2,calc}$. The color indicates the impact velocity. The gray shaded region is where $M_{CO2,calc}$ is not reliable because of low spatial resolution (see text).

## 4. Discussion

The similarity between $M_{CO2,exp}$ and $M_{CO2,calc}$ at $Y_{lim}$ = 1.3–2.6 GPa suggests that our two-step model can reproduce impact devolatilization of natural calcite samples in the laboratory, although a larger $Y_{lim}$ (2–4 times than the fiducial value) compared with the fiducial value (0.65 GPa) yields the best-fit results. The difference in $Y_{lim}$ is best explained by the dynamic failure mechanism [e.g., Ramesh et al., 2015], as noted by Melosh and Ivanov [2018]. Given that the strain rate $\dot{\varepsilon}$ in impact experiments conducted by Kurosawa et al. (2012) ($\dot{\varepsilon} \sim t_s^{-1} \sim 10^6$ s$^{-1}$) is much higher than that usually achieved in static failure experiments ($\dot{\varepsilon} < 10^{-4}$ s$^{-1}$ [e.g., Hacker and Kirby, 1993]), the brittle solids become much stronger according to the power law for the strain rate ($Y \propto \dot{\varepsilon}^{2/3}$, where $Y$ is compressive yield strength) [Ramesh et al., 2015]. Given that the reference strength of the limestone listed in Table 1 of Ramesh et al. [2015] was extrapolated to a strain rate of $10^6$ s$^{-1}$ with the power law, the compressive yield strength of the limestone was estimated to be ca. 10 GPa, which is 4–8 times the best-fit value of $Y_{\text{lim}}$, if the extrapolation is valid. However, intense shock compression does not only cause rock failure, but also leads to irreversible shock heating, resulting in thermal softening. The thermal softening limits the degree of post-shock heating due to plastic deformation [Pierazzo and Melosh, 2000; Kurosawa and Genda, 2018]. Based on the above discussion, we propose that the "effective" $Y_{\text{lim}}$ for calcite, including the effects of mechanical hardening at a high strain rate and thermal softening, is 1.3–2.6 GPa under the conditions of the previous experiment [Kurosawa et al., 2012]. Given the strain rate in the case of natural impact events would be much lower ($\dot{\varepsilon} \sim t_s^{-1} \sim 1$ s$^{-1}$; e.g., for a collision with an impactor with a diameter of 5 km at 5 km s$^{-1}$) than the case of laboratory experiments, the "effective" $Y_{\text{lim}}$ would be closer to the fiducial value of $Y_{\text{lim}}$ than shown in this study. The effect of thermal softening on $M_{\text{CO2,calc}}$ is addressed in the Supplementary Information. In the case without thermal softening, $M_{\text{CO2,calc}}$ is reduced by a factor of 2–4.

In this study, we do not focus on projectile behavior. Projectiles are expected to also suffer post shock heating. Kenkmann et al. (2013) conducted an impact experiment with a steel projectile and a porous sandstone at 5.3 km s$^{-1}$. They reported that unexpected melting of a steel projectile occurs at ~70 GPa of shock pressure. The shock pressure is much lower than the shock pressure required for incipient melting of iron (~200 GPa, estimated by using the entropy required for incipient melting of iron [Pierazzo et al., 1997] and the entropy of the iron Hugoniot [Kraus et al., 2015]). It is presumed that the heat source is plastic deformation of the projectile and/or thermal conduction from the strongly heated target due to porosity compaction [Kenkmann et al., 2013].

Pierazzo and Melosh (2000) argued that the self-limiting nature of strength-induced heating prevents the enhancement of the degree of impact vaporization. In contrast, impact devolatilization is more significant than previously thought, as shown by the present study, because the required temperatures for incipient devolatilization $T_{\text{id}}$ of volatile-bearing rocks are generally lower than the typical melting temperature $T_{\text{melt}}$ where the yield strength becomes zero. Since thermal metamorphism, such as changes in reflectance spectra (ca. 800 K) and atomic diffusion (ca. 1,000 K), also proceeds at similar temperatures to $T_{\text{id}}$ [e.g., Hiroi et al., 1996; Nakato et al., 2008; Cohen, 2013], thermal metamorphism recorded in meteorites [e.g., Stöffler et al., 2018] could have been caused by impacts at much lower velocities than previously thought. It should be noted that the present results for macro blocks of calcite cannot be applied directly to parent bodies of chondrites. We used calcium carbonate as a typical geologic material: an elasto-plastic medium with a certain strength. As demonstrated in Fig. 1, the pressure–strength behavior of calcite can be expressed by the "ROCK" model. This implies that the results for calcite are able to be applied to actual chondritic materials, although in-depth and careful analyses using actual chondritic materials are required for confirmation of this finding.

Schultz (1996) used oblique impact experiments to show that shear heating in oblique impacts enhances the total vapor production. Wakita et al. [2019] has also shown numerically with the three-dimensional version of iSALE that oblique impacts lead to more efficient post-shock heating due to plastic deformation than vertical impacts. However, the enhancement of post-shock heating by oblique impacts should be tested in future experimental studies.

Finally, we would like to mention about the importance of the choices of the parameters and their accuracy for strength models. As demonstrated in this study, the implementation of strength models to hydrocode modeling is crucial to estimate the heated mass accurately especially at low-velocity impacts. Our numerical model clearly demonstrates that the accuracy of the input parameters strongly affects the heated mass. For example, $M_{CO2,calc}$ at $v_{imp}$ = 4.14 km s$^{-1}$ and $Y_{lim}$ = 2.6 GPa becomes ~30 times larger than that at the fiducial value. Thus, subtle changes can yield results that differ by more than one order of magnitude in terms of impact devolatilization.

5. Conclusions

Shock-induced thermal metamorphism in meteorites records the ancient impact environment in the Solar System. If we could accurately understand impact heating, then it would be possible to reveal the impact history of the Solar System. In this study, we investigated the significance of the recently-realized heat source in natural samples, that is, the post-shock heating due to plastic deformation. Impact simulations were compared with a previous experiment on the impact devolatilization of calcite marble to explore the influence of plastic deformation on heating. The simulation reproduces the devolatilization behavior of calcite with increasing impact velocity, which demonstrates that heating due to plastic deformation is also significant in natural samples. This finding possibly suggests that the impact environments inferred from the thermal metamorphism of meteorites are revised to be gentler than previously expected.

**Acknowledgments**

We thank the developers of iSALE, including G. Collins, K. Wünnemann, B. Ivanov, J. Melosh, and D. Elbeshausen. We used pySALEplot to analyze the output file of iSALE and to make Fig. 3. We also thank Tom Davison for the development of pySALEPlot. We appreciate Boris Ivanov for providing the modified parameter set of the ANEOS for calcite. We thank two anonymous referees for their careful reviews that helped greatly improve the manuscript, and Andrew J. Dombard for handling the manuscript as an editor. K.K. is supported by JSPS KAKENHI grants JP17H01176, JP17H01175, JP17K18812, JP18H04464, and JP19H00726. H.G. and K.K. were supported by JSPS KAKENHI grant JP17H02990. H.G. is supported by MEXT KAKENHI Grant JP17H06457. A part of numerical computations was carried out on the general-purpose PC cluster at Center for Computational Astrophysics, National Astronomical Observatory of Japan. The iSALE shock physics code is not fully open-source, but is distributed on a case-by-case basis to academic users in the impact community for non-commercial use. A description of the application requirements can be found at the iSALE website (http://www.isale-code.de/redmine/projects/isale/wiki/Terms_of_use). The M-ANEOS package has been available from Thompson SL, Lauson HS, Melosh HJ, Collins GS and Stewart, ST, M-ANEOS: A Semi-Analytical Equation of State Code, Zenodo, http://doi.org/10.5281/zenodo.3525030. The lists of input parameters of the iSALE computation can be found in the Supplementary Information. The NASA CEA code could be used via the website (https://www.grc.nasa.gov/www/CEAWeb/).

Takata, T. and T. J. Ahrens (1994), Numerical simulation of impact cratering at Chicxulub and the possible causes of KT catastrophe, in *New Developments Regarding the KT Event and Other Catastrophes in Earth History*, Contrib.825, pp. 125-126, Lunar and Planetary Institute, Houston, Tex.

Thompson, S. L., and H. S. Lauson (1972), Improvements in the Chart-D radiation hydrodynamic code III: Revised analytical equation of state, pp. *SC-RR-71 0714* **119** pp., Sandia Laboratories, Albuquerque, NM.

Tillotson, J. H. (1962), Metallic equations of state for hypervelocity impact, *Technical Report GA-3216*, General Atomic Report.

Wakita, S., H. Genda, K. Kurosawa, and T. M. Davison (2019), Enhancement of impact heating in pressure-strengthened rocks in oblique impacts, *Geophysical Research Letters*, **46**, 13678-13686, https://doi.org/10.1029/2019GL085174.

Wünnemann, K., G. S. Collins, and H. J. Melosh (2006), A strain-based porosity model for use in hydrocode simulations of impacts and implications for transient crater growth in porous targets, *Icarus*, **180**, 514–527.

Wünnemann, K., G. S. Collins, and G. R. Osinski (2008), Numerical modelling of impact melt production in porous rocks, *Earth and Planetary Science Letters*, **269**, 530-539.
18

Supporting Information

**This PDF file includes:**

Supplementary text S1 to S7
Figures S1 to S4
Tables S1 to S4
SI References

**S1. Theoretical background to post-shock heating due to plastic deformation**

The temperature rise $\Delta T$ due to the plastic deformation of shock-comminuted rocks is characterized by the following relationships [Kurosawa and Genda, 2018; Melosh and Ivanov, 2018]:

$$e_{\text{strength}} \sim \varepsilon Y_d / \rho \, (\text{J kg}^{-1}), \qquad (1)$$

$$Y_d = \min(Y_{\text{coh}} + \mu P, Y_{\text{lim}}) \, (\text{Pa}) \qquad (2)$$

$$\Delta T = e_{\text{strength}}/C_p \, (\text{K}), \qquad (3)$$

where $e_{\text{strength}}$, $\varepsilon$, $Y_d$, $\rho$, $Y_{\text{coh}}$, $\mu$, $P$, $Y_{\text{limit}}$, and $C_p$ are the plastic work, volumetric strain, yield strength of shock-comminuted rocks, density, cohesion, frictional coefficient, pressure of the distorted materials in the shock-driven flow field, von Mises limit, and isobaric specific heat, respectively. Equation (2) corresponds to the well-known Mohr–Coulomb friction law, and is considered to be a good approximation of the strength behavior of shock-comminuted materials (i.e., granular materials) [e.g., Drucker and Prager, 1952; Collins et al., 2004]. Equations (1)–(3) suggest that heating due to plastic deformation is expected to be efficient at the location with a large $\varepsilon$ that is close to unity. Thus, heating becomes significant around the impact point, because of the large shear deformation (i.e., a large $\varepsilon$ is expected). In addition, the yield strength of rocky materials also depends on the temperature of the distorting media [Ohnaka, 1995]. At the melting point, $Y_d$ becomes zero. Thus, the degree of post-shock heating due to plastic deformation becomes negligible at relatively high-speed collisions of >10 km s$^{-1}$, because materials with large $\varepsilon$ would be melted by such hypervelocity impacts.

**S2. Previous impact experiments**

Kurosawa et al. [2012] conducted a series of impact experiments using block targets made from Carrara marble. Hecker and Kirby [1993] also used Carrara marble in their static deformation experiment. The Carrara marble is composed of randomly oriented calcite ($CaCO_3$) crystals with a negligible porosity. Kurosawa et al. [2012] developed a new experimental technique using two-stage light gas guns to investigate shock-induced devolatilization from unconfined targets in an open system, with a minimal risk of chemical contamination. The geometry of the impacts is the same as for natural impact event (i.e., a spherical projectile collides onto a semi-infinite flat plane). Thus, the pressure and temperature paths of shocked calcite in the laboratory are expected to be the same as for natural impact events. Consequently, the experimental data is suitable for comparing with the model results. The details of the method and procedures are described in Kurosawa et al. [2012]. The masses of the $CO_2$ produced $M_{CO2,exp}$ at given impact velocities are listed in Table S1. The uncertainty in the measurements is ±10%, which derives from the calibration procedure used by Kurosawa et al. (2012).

**Table S1.** Experimental results of Kurosawa et al. [2012].

| Impact velocity (km s$^{-1}$) | $CO_2$ mass (normalized by the projectile mass) |
|---|---|
| 1.88 | 3.1 × 10$^{-4}$ |
| 2.38 | 4.5 × 10$^{-3}$ |
| 3.28 | 1.4 × 10$^{-2}$ |
| 4.14 | 2.2 × 10$^{-2}$ |
| 5.03 | 7.5 × 10$^{-2}$ |
| 6.06 | 1.2 × 10$^{-1}$ |
| 6.66 | 1.6 × 10$^{-1}$ |

**S3. Model input parameters and calculation settings**

The model input parameters for calcite are listed in Table S2. A detailed description of each parameter used in iSALE can be found in Collins et al. [2016]. The material strength of the $Al_2O_3$ projectile was not included in the calculations. Since the Tillotson parameters for $Al_2O_3$ are not available in the literature, we determined the parameters listed in Table S3 from the experimental data of Marsh [1980]. Figure S1 shows the calculated Hugoniot curve on a density–pressure plane along with the experimental data. The horizontal dotted lines correspond to the shock pressures achieved in the experiment by Kurosawa et al. (2012). The parameter set listed in Table S3 yields a good fit to the Hugoniot curve under the experimental conditions.

We describe the iSALE model setup in Table S4. The full description of each parameter can also be found in the iSALE manual [Collins et al., 2016]. Gravity has a negligible effect because impact devolatilization occurs very early in the impact process. The calculations were run until the time when the calculated mass of $CO_2$ produced became constant (see Section 2.2 in the main text). We conducted a total of 56 runs with 7 different $Y_{lim}$ values (0.65, 0.85, 1.0, 1.3, 1.6, 2.0, and 2.6 GPa) and no material strength (i.e., purely hydrodynamic conditions). Although we introduced high-velocity and low-density cutoffs into the calculation to stabilize the numerical integration, the effects of the artificial cutoffs can be neglected because of the sufficiently loose values. Given that our numerical model does not include gravity and size-dependent strength, the calculation results do not depend on the actual grid size.

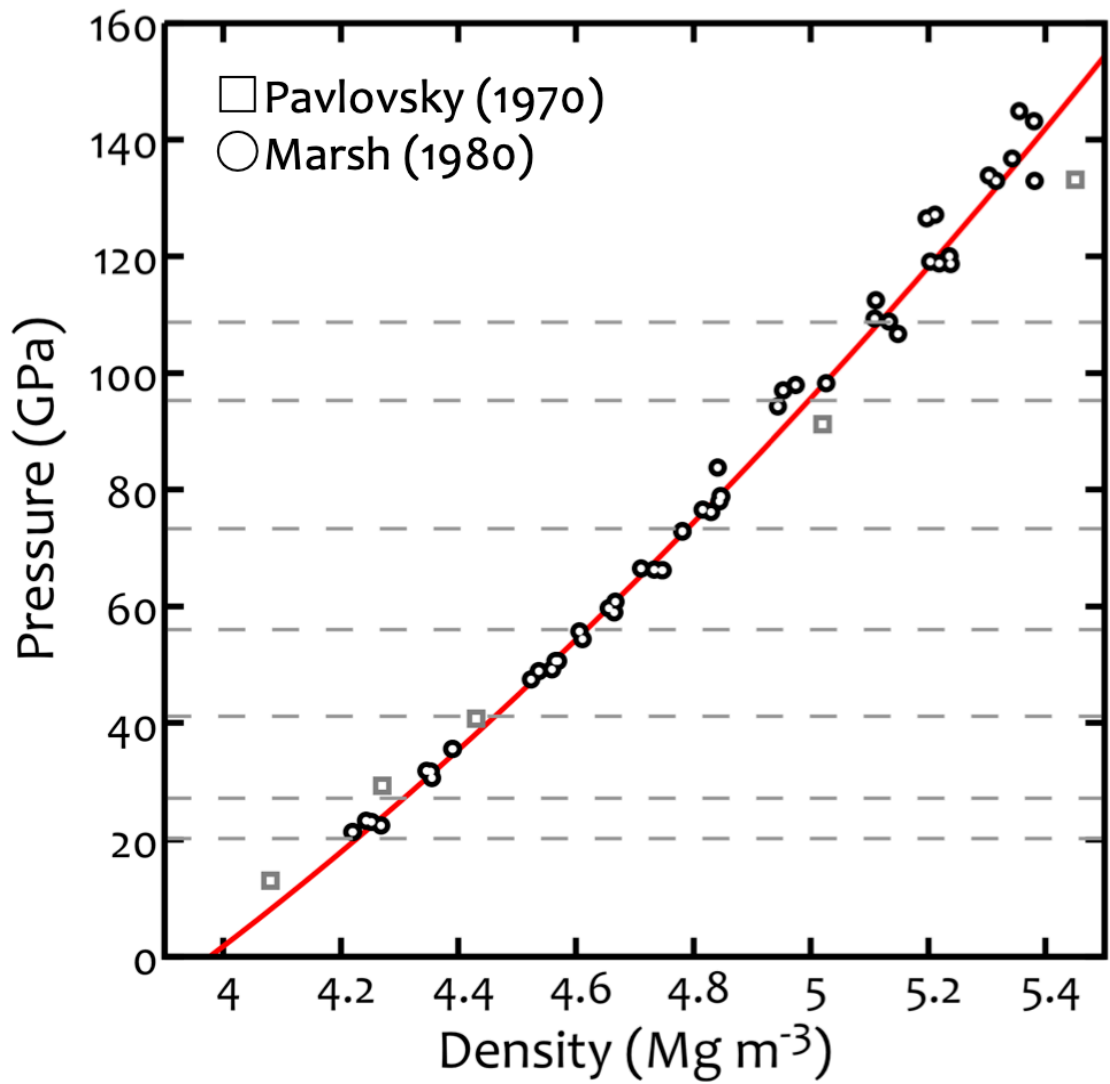

**Figure S1.** The Hugoniot curve of $Al_2O_3$ on a density-pressure plane calculated by the Tillotson EOS with the parameters listed in Table S3 along the experimental data by Pavlocsky (1970) and Marsh (1980). The horizontal lines correspond to the peak shock pressures at the impact points.

**Table S2.** Model input parameters for calcite.

| EOS type | ANEOS[a] |
|---|---|
| Strength model | Rock |
| Poisson ratio | 0.3 |
| Melting temperature (K) | 1690[b] |
| Thermal softening coefficient | 1.2[c] |
| Simon parameter $a$ (GPa) | 24.3[d] |
| Simon parameter $c$ | 2.27[d] |
| Cohesion (undamaged) (MPa), $Y_{coh,i}$ | 50 |
| Cohesion (damaged) (MPa), $Y_{coh}$ | 0.1 |
| Internal friction (undamaged), $\mu_{int}$ | 2 |
| Internal friction (damaged), $\mu_{dam}$ | 0.4 |
| Limiting strength (GPa), $Y_{limit}$ | 0.65, 0.85, 1.0, 1.3, 1.6, 2.0, 2.6 |
| Minimum failure strain | $10^{-4}$ [c] |
| Constant for the damage model | $10^{-11}$ [c] |
| Threshold pressure for the damage model (MPa) | 300 [c] |

a. The parameter set for calcite is basically the same as that of Pierazzo et al. [1998]. The bulk sound speed $C_s$ and Grüneisen parameter $\Gamma_0$ at the reference state were slightly modified. The modified values of $C_s$ and $\Gamma_0$ are 3.35 km s$^{-1}$ and 1.6, respectively [Ivanov, personal communication, 2018]. The reference value of entropy $S_0$ is adjusted to the published thermodynamic data ($S_0 = 917$ J K$^{-1}$ kg$^{-1}$) [Jacobs et al., 1981].
b. The lowest melting temperature on a pressure-temperature plane (3 MPa) from Kerley [1989] was used for the melting temperature under the reference state.
c. Typical values for minerals were used [Ivanov et al., 1997; Collins et al., 2004].
d. The solidus pertaining to calcite from Kerley [1989] was fitted with the Simon equation [e.g., Wünnemann et al., 2008].

**Table S3.** Tillotson parameters for $Al_2O_3$. Detailed descriptions of the parameters can be found in previous studies [e.g., Tillotson, 1962; Melosh, 1989].

| Reference density $\rho_0$ (kg m$^{-3}$) | 3,977 |
|---|---|
| Bulk modulus $A$ (GPa) | 303[a] |
| Tillotson $B$ constant (GPa) | 142[b] |
| Fitting constant $E_0$ (MJ/kg) | 2.0[c] |
| Tillotson $a$ constant | 0.35[c] |
| Tillotson $b$ constant | 0.585[c] |
| Tillotson $\alpha$ constant | 5[d] |
| Tillotson $\beta$ constant | 5[d] |
| Specific internal energy for incipient vaporization $E_{iv}$ (MJ/kg) | 4.1[e] |
| Specific internal energy for incipient vaporization $E_{cv}$ (MJ/kg) | 18[f] |

a. The bulk modulus $A$ was calculated as $A = \rho_0 C_s^2$, where $C_s$ is the bulk sound speed [e.g., Melosh, 1989]. Here, $C_s$ was estimated to be 8.7 km s$^{-1}$ from the data compilation by Marsh (1980)
b. The Tillotson $B$ constant was estimated from the frequently used empirical relationships using the Grüneisen coefficient $\Gamma_0$ at the reference state and the bulk modulus (i.e., $B = A\Gamma_0/2$) [e.g., Melosh, 1989].
c. These values were determined by fitting the shock Hugoniot data for $Al_2O_3$ from Marsh [1980].
d. The Tillotson $b$ constant was calculated by the frequently used empirical relationships between the shock Hugoniot constant $s$, $\Gamma_0$, and Tillotson $a$ constant (i.e., $b = \Gamma_0 - a$ and $\Gamma_0 = 2s - 1$) [e.g., Melosh, 1989].
e. Typical values were used [e.g., Melosh, 1989].
f. $E_{iv}$ was calculated from the isobaric specific heat at the high-temperature limit $C_p = 1.27$ kJ K$^{-1}$ kg$^{-1}$ [Xu et al., 1995] and boiling temperature $T_{boil}$ of $Al_2O_3 = 3250$ K [e.g., Haynes, 2010] as $E_{iv} = C_p T_{boil}$. We assumed that the isochoric specific heat $C_v$ is well approximated by $C_p$.

g. $E_{cv}$ can be calculated as $E_{cv} = E_{iv} + L_{vap}$, where $L_{vap}$ is the latent heat of vaporization. The latent heat $L_{vap}$ for $Al_2O_3$, however, is not available in the literature. Thus, $E_{cv}$ was set to a value typical of oxides [Melosh, 1989].

**Table S4.** Calculation settings.

| | |
|---|---|
| Computational geometry | Cylindrical coordinates |
| Number of computational cells in the $R$ direction | 2000 |
| Number of computational cells in the $Z$ direction | 3000 |
| Cells per projectile radius (CPPR) | 200 |
| Artificial viscosity $a_1$ | 0.24 |
| Artificial viscosity $a_2$ | 1.2 |
| Impact velocity (km s$^{-1}$) | Same as for the impact experiments listed in Table S1 |
| High-speed cutoff | Double the impact velocity |
| Low-density cutoff (kg m$^{-3}$) | 1 |

**S4. Purely hydrodynamic calculations**

We also conducted the iSALE computation without material strength, as mentioned in the main text. Figure S2 shows the same as Fig. 3 in the main text, except that the results at an impact velocity of 4.14 km s$^{-1}$ without material strength are shown. The entropy gains of the shocked calcite are well below 0.8 kJ K$^{-1}$ kg$^{-1}$, even near the crater wall as shown in panel (a). We confirmed that the thermodynamic paths of the selected tracers mostly follow the isentropes from the Hugoniot curve on an entropy–pressure plane, as conventionally assumed in a number of previous studies [e.g., Ahrens and O'Keefe, 1972]. The tracers having the entropy required for incipient devolatilization were not recorded in this simulation, whereas $CO_2$ production from shocked calcite was observed in the impact experiments [Kurosawa et al., 2012].

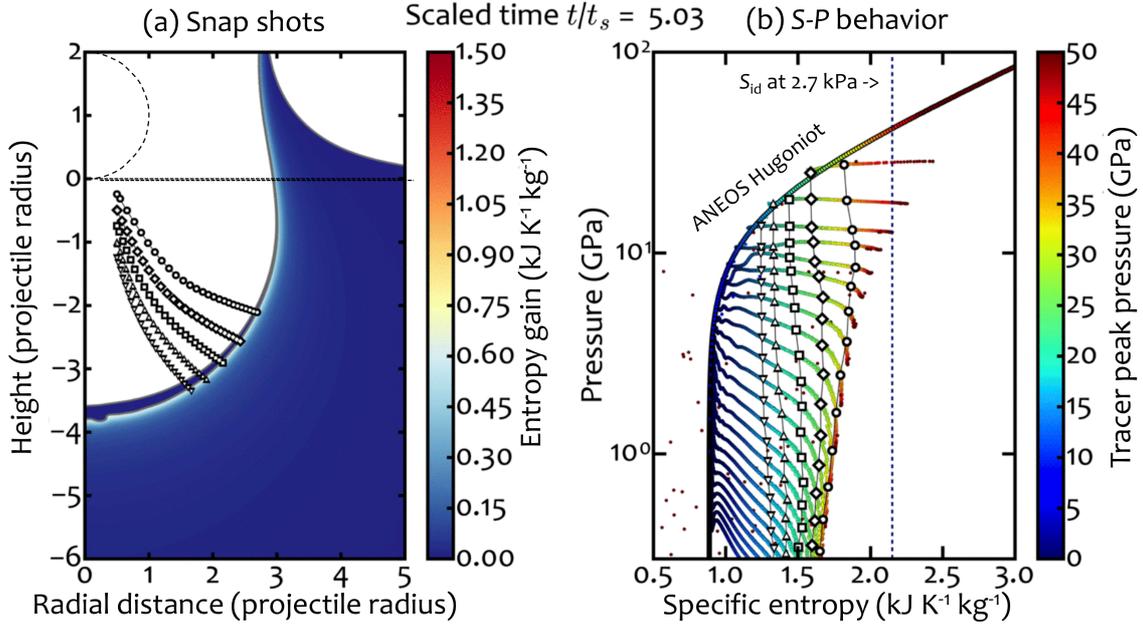

**Figure S2.** Same as Fig. 3 in the main text, but calculated without material strength.

### S5. Resolution effects on the mass of $CO_2$ produced

It is well known that a low spatial resolution in numerical simulations sometimes yields numerical artifacts. Thus, we must address the resolution effects on the mass of $CO_2$ produced in the simulation of $M_{CO2,calc}$. Another series of simulations was conducted by changing the number of cells for the projectile radius $n_{CPPR}$ from 13 to 200 at the limiting strength $Y_{lim}$ of 1.3 GPa. Figure S3 shows the ratio of $M_{CO2,calc}$ to $M_{CO2,exp}$ as a function of $n_{CPPR}^{-1}$, which indicates that the $M_{CO2,calc}$ values are inversely proportional to $n_{CPPR}^{-1}$ when $n_{CPPR} \geq 50$. This allows us to estimate $M_{CO2,calc}$ at infinite spatial resolution ($n_{CPPR} = \infty$). The differences in the ratios between $n_{CPPR} = 200$ and $\infty$ for all the shots shown here are within a factor of 1.25. The shot at the lowest $v_{imp}$ (1.88 km s$^{-1}$) yields a too large amount of $CO_2$ (Fig. 4), as compared with $M_{CO2,exp}$. This overshoot is possibly caused by a low spatial resolution. The $M_{CO2,exp}$ value of this shot (0.03 wt.% of the projectile mass) corresponds to 10% of the mass of the calcite cylinder below the impactor footprint, with a one cell thickness even in the case of $n_{CPPR} = 200$. Although we introduced artificial viscosity into the simulation, it is difficult to prevent some degree of temperature overshoot within a few cells around the contact boundary between the projectile and target. Thus, the overshoot in $M_{CO2,exp}$ at the lowest $v_{imp}$ of 1.88 km s$^{-1}$ is likely to be a numerical artifact near the contact boundary.

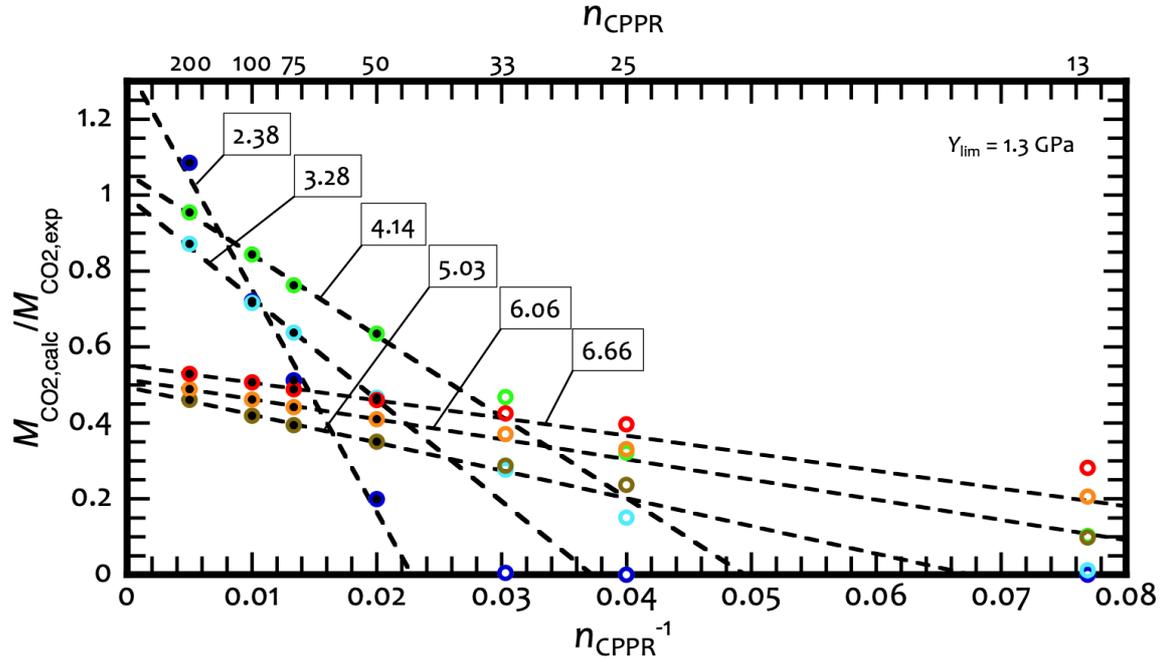

**Figure S3.** Resolution effect on the ratio of the mass of $CO_2$ produced in the simulation. The impact velocities are shown in units of km s$^{-1}$. The dashed lines are the best-fit linear trends for data with $n_{CPPR}^{-1} \geq 50$. The data (symbols) used for the fitting are infilled with a black color.

## S6. Uncertainties in the numerical model

Here, we discuss the possible uncertainties associated with the numerical model. The resolution effects, as mentioned in the previous section, constitute one of the major uncertainties. The thermodynamic behavior of shocked calcite in iSALE has not been tested by comparing impact experiments. However, the shock pressure distribution given by iSALE has been tested in a shock recovery experiment by using single-crystal San Carlos olivine [Nagaki et al., 2016]. Those authors also used ANEOS and the ROCK model, as in the present study, but with parameters for dunite. They reported that the texture of the recovered shocked olivine is consistent with the calculated shock pressure distribution given by iSALE.

Our two-step method depends on the accuracy of the thermodynamic data for the Ca–C–O systems, which are used to calculated the decomposition boundaries $S_{id}$ and $S_{cd}$ at the reference pressure, as mentioned in the main text. Thus, if the data are revised, the calculated total emissions of $CO_2$ will be also revised.

## S7. The effect of thermal softening on the total $CO_2$ production

In the discussion section of the main text, we examine the possibility that rate-induced strengthening is compensated for by thermal softening. Here, we present the iSALE results for the case without the Ohnaka thermal softening model. Figure S4 shows a comparison of total $CO_2$ production among the impact experiment, iSALE with thermal softening, and iSALE without thermal softening. The impact velocities were the same as those used in the impact experiment. The limiting strength $Y_{lim}$ was fixed at 0.65 GPa. We found that thermal softening reduces the total $CO_2$ emissions by a factor of 4.4 at 4.14 km s$^{-1}$ and by 1.9 at 6.66 km s$^{-1}$, suggesting that thermal softening should occur in the impact experiment. In other words, the scaling law for dynamic failure constructed under room temperature does not apply to shock-induced deformation in the laboratory. The difference in the total $CO_2$ emissions could be compensated for by using a somewhat larger value of limiting strength, as shown in Figure 4 in the main text.

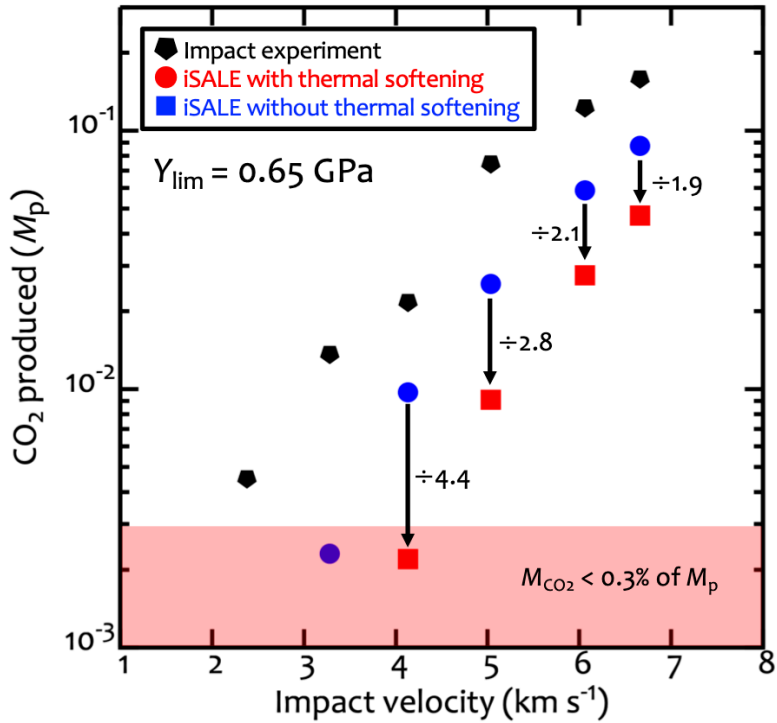

**Figure S4.** The total $CO_2$ production as a function of impact velocity. The results from the impact experiment by Kurosawa et al. (2012) (black pentagons), the iSALE with thermal softening (red squares), and the iSALE without thermal softening (blue circles). The red shaded region is where $M_{CO2,calc}$ is not reliable because of low spatial resolution as mentioned in the main text.